\documentclass[aps,prl,letterpaper,twocolumn,showpacs,superscriptaddress,floatfix,longbibliography]{revtex4}

\usepackage[english]{babel}
\usepackage{amsmath}
\usepackage{color}
\usepackage{epsfig}
\usepackage{graphicx}
\usepackage{bm}
\usepackage{times,amsmath,amssymb}
\usepackage{subfigure}
\usepackage{amsfonts}
\usepackage{amssymb}
\usepackage{amsmath}
\usepackage{color}
\usepackage{mathtools}
\usepackage{empheq}

\global\long\def\bra#1{\langle #1 |}
\global\long\def\ket#1{| #1 \rangle }

\global\long\def\al{\alpha}
\global\long\def\no{\nonumber}
\global\long\def\be{\beta}
\global\long\def\ga{\gamma}

\global\long\def\De{\Delta}

\global\long\def\th{\theta}

\global\long\def\si{\sigma}

\newcommand{\nc}{\newcommand}
\nc{\ir}{\mathrm{i}}
\nc{\dd}{\mathrm{d}}   
\nc{\eE}{\mathsf{e}}
\nc{\tM}{{\widetilde{M}}}

\def\bee{\begin{eqnarray}}
\def\eee{\end{eqnarray}}
\def\ir{{\mathrm i}}
\def\rP{{\mathrm p}}

\def\eE{\mathsf{e}}

\def\sh{{\sinh }}
\def\ch{{\cosh }}

\begin{document}

\title{Chiral coordinate Bethe ansatz for phantom eigenstates in the open XXZ
  spin-$\frac12$ chain }

\author{Xin Zhang}
\affiliation{Beijing National Laboratory for Condensed Matter Physics, Institute of Physics, Chinese Academy of Sciences, Beijing 100190, China}
\affiliation{Department of Physics,University of Wuppertal, Gaussstra\ss e 20, 42119 Wuppertal, Germany}

\author{Andreas Kl\"umper}
\affiliation{Department of Physics,University of Wuppertal, Gaussstra\ss e 20, 42119 Wuppertal, Germany}

\author{Vladislav Popkov}
\affiliation{Faculty of Mathematics and Physics, University of Ljubljana, Jadranska 19, SI-1000 Ljubljana, Slovenia}
\affiliation{Department of Physics,University of Wuppertal, Gaussstra\ss e 20, 42119 Wuppertal, Germany}

\begin{abstract}

  We construct the coordinate Bethe ansatz for all eigenstates of the open
spin-$\frac12$ XXZ chain that fulfill the phantom roots criterion (PRC). Under
the PRC, the Hilbert space splits into two invariant subspaces and there are two sets of
homogeneous Bethe ansatz equations (BAE) to characterize the subspaces
in each case. We propose two sets of vectors with chiral shocks to span the
invariant subspaces and expand the corresponding eigenstates. All the vectors
are factorized and have symmetrical and simple structures. Using several
simple cases as examples, we present the core elements of our generalized
coordinate Bethe ansatz method. The eigenstates are expanded in our generating
set and show clear chirality and certain symmetry properties. The bulk
scattering matrices, the reflection matrices on the two boundaries and the BAE
are obtained, which demonstrates the agreement with other approaches.  Some
hypotheses are formulated for the generalization of our approach.
\end{abstract}

\maketitle


\section{Introduction}
Quantum integrable systems \cite{Baxter,Sklyanin,GaudinBook} play important
roles in various fields such as low-dimensional condensed matter physics,
quantum field theory, statistical physics and Yang–Mills theory. Many methods
have been developed for the analysis of integrable systems. Among them, the
two most classic ones are the coordinate Bethe ansatz and the algebraic Bethe
ansatz (ABA).  The usage of the conventional coordinate Bethe ansatz and ABA
has so far been restricted to one-dimensional integrable systems with $U(1)$
symmetry that guarantees the existence of some obvious reference
states. For integrable systems without $U(1)$ symmetry, there are no
  obvious reference states and the conventional BA fails. Several methods
including Baxter's $T$-$Q$ relation \cite{Baxter} and Sklyanin's separation of
variables (SoV) method \cite{Sklyanin1995} have been developed to approach
this remarkable problem.


In this paper we focus on the XXZ spin-$\frac12$ chain with open
boundaries. The non-diagonal boundary fields break the $U(1)$ symmetry which
makes the problem of constructing Bethe vectors rather unusual.  It was proved
in \cite{OffDiagonal03,Nepomechie2003,Rafael2003} for the boundary
parameters obeying a certain constraint, that the modified ABA can be applied and
homogeneous conventional $T$-$Q$ relations exist. The eigenvalue problem of
the open XXZ spin chain with generic integrable boundary conditions was first
solved via the off-diagonal Bethe ansatz (ODBA) method
\cite{Cao2013off,OffDiagonal}. The Bethe-type eigenstates were then retrieved
in \cite{Zhang2015} based on the ODBA solution and a convenient SoV basis
\cite{Niccoli2012,Faldella2014,Kitanine2014}. Although the analytical form of
the Bethe state with generic or constrained boundaries has been given, little
is known about their inner structure.
 
In our recent papers \cite{PhantomLong,PhantomShort}, we studied the
eigenstates of the open XXZ chain under the phantom roots criterion (PRC). The
PRC is equivalent to the constrained boundary condition proposed in
\cite{OffDiagonal03,Nepomechie2003}. The PRC restricts the system parameters
to a set of manifolds parameterized by an integer number $M$ but does not
introduce any obvious symmetry like $U(1)$ symmetry. Under the PRC, the
Hilbert space splits into two invariant subspaces whose dimensions are
determined by the integer $M$. Two sets of factorized chiral states are
selected here to span the subspaces respectively. In
\cite{PhantomLong,PhantomShort} we constructed the phantom Bethe states in
some simple cases and analyzed their properties such as the chirality and the
corresponding spin current.

In this paper, the coordinate Bethe Ansatz method is generalized in full
detail.  Here we report on a formulation of ``generalized" chiral coordinate
Bethe ansatz (CCBA) in an open XXZ spin chain with non-diagonal boundary
fields.  We do this on the example of the system satisfying the PRC.

Our approach inherits the core ideas of conventional coordinate Bethe ansatz
method and gives novel results. The novelty is two-fold: (\romannumeral1) we
find that it is appropriate to use the basis vectors with chiral shocks,
instead of the usual conventional computational basis; for this reason we
  also call it a chiral Bethe ansatz  (\romannumeral2) it turns out to be
appropriate to enlarge the basis into a symmetric one by including
linearly-dependent ``auxiliary" vectors.
  
The paper is organized as follows. First, we introduce the open XXZ
spin-$\frac12$ chain under the phantom roots conditions. Two symmetrically
enlarged sets of vectors are then constructed based on which we can expand the
phantom eigenstates of the Hamiltonian. Next, we demonstrate how the chiral
coordinate Bethe ansatz works in terms of these vectors for the $M=0,1,2$ cases
and generalize our method to the arbitrary $M$ case. In the last part of the main
text, we specifically study the spin helix eigenstates. Some necessary proofs
are given in the Appendices.


\section{The open XXZ model under Phantom Roots conditions}

We study the spin-$\frac12$ XXZ chain with open boundary conditions

\begin{align}
&H= \sum_{n=1}^{N-1}  h_{n,n+1}+ h_1+h_N,\label{eq:XXZopen}
\end{align}
where 
\bee 
&&h_{n,n+1}\!=\!\sigma_n^x\sigma_{n+1}^x\!+\!\sigma_n^y\sigma_{n+1}^y\!+\!\ch\eta\sigma_n^z\sigma_{n+1}^z\!-\!\ch\eta\,I,\\[2pt]
&&h_1=\frac{\sh\eta}{\sh(\al_-)\ch(\be_-)}(\ch(\theta_-)\sigma_1^x+\ir\,\sh(\theta_-)\sigma_1^y\no\\
&&\hspace{0.8cm}+\ch(\al_-)\sh(\be_-)\sigma_1^z),\\[2pt]
&&h_N=\frac{\sh\eta}{\sh(\al_+)\ch(\be_+)}(\ch(\theta_+)\sigma_N^x+\ir\,\sh(\theta_+)\sigma_N^y\no\\
&&\hspace{0.8cm}-\ch(\al_+)\sh(\be_+)\sigma_N^z),
\eee 
and $\al_\pm$, $\be_\pm$, $\th_{\pm}$ are boundary parameters. We parameterize the anisotropy parameter of the exchange interaction as
$\De\equiv\cosh \eta\equiv\cos \gamma$ with $\eta=\ir\gamma$.

This model is one of the most famous integrable systems
\cite{Baxter,FaddeevTakhtajan,SklyaninFaddeevTakhtajan,Sklyanin} without
$U(1)$ symmetry. The exact solutions of this model have been given by the ODBA
method \cite{Cao2013off,OffDiagonal}. A set of inhomogeneous Bethe ansatz
equations (BAE) with at least $N$ Bethe roots were constructed
\cite{Cao2013off,OffDiagonal,Zhang2015} to solve the eigenvalue problem and
the Bethe-type eigenstates were then retrieved \cite{OffDiagonal,Zhang2015}
based on the ODBA solution.

An interesting observation is that some Bethe roots in the original
inhomogeneous BAE can be chosen ``phantom'', i.e.~with infinite value of
the root and hence not contributing to the energy, under some specific
conditions like
\begin{align}
&(N-2M-1)\eta \no\\
& = \al_{-}+ \be_{-}
+ \al_{+}+ \be_{+} + \th_{-} -\th_{+}\,\,\mod 2\pi \ir,  \label{ConditionPhantomPlus}
\end{align}
where $M$ is an integer ranging from 0 to $N-1$.  Under the phantom Bethe
roots criterion (PRC) (\ref{ConditionPhantomPlus}), the inhomogeneous BAE can
reduce to homogeneous ones with $M$ or $\tM=N-1-M$ preserved finite Bethe roots
\cite{OffDiagonal,Nepomechie2003,Rafael2003,OffDiagonal03} and the Hilbert
space splits into two invariant subspaces $G_M^+$ and $G_M^-$, whose
dimensions are determined by the integer $M$
\cite{PhantomLong,PhantomShort}. The PRC also serve as the compatibility
condition of the modified ABA method \cite{OffDiagonal03,Belliard}.

Under the constraint (\ref{ConditionPhantomPlus}), the hermiticity of
Hamiltonian (\ref{eq:XXZopen}) requires in the case $|\Delta|<1$ (the easy
plane regime)
\bee 
&&{\rm Re[\al_{\pm}]=Re[\theta_{\pm}]= Re[\eta]=0},\no\\
&&\rm Im[\be_{\pm}]=0\,\,{\mbox and }\,\,\be_+=-\be_-,\label{hermiticity_1}
\eee
and in the case $\Delta>1$ (the easy axis regime)
\bee 
&&{\rm Im[\al_{\pm}]=Im[\be_{\pm}]= Im[\eta]=0},\no\\
&&\rm Re[\theta_{\pm}]=0\,\,{\mbox and }\,\,\theta_+=\theta_-\,\,mod \,\,2\ir\pi.\label{hermiticity_2}
\eee 


In the following we show that the two sets of homogeneous BAE 
correspond to two  invariant subspaces $G_M^+$ and $G_M^-$ respectively,
and their solutions constitute the complete set of eigenstates and eigenvalues 
under the criterion (\ref{ConditionPhantomPlus}). In addition, we construct explicit phantom Bethe vectors via a chiral coordinate Bethe ansatz, see below.

\section{Addition of extra auxiliary vectors to the bases of $G_M^{\pm}$.}



Here we explain a perhaps most important and subtle feature of the chiral
coordinate Bethe ansatz for open systems with non-diagonal boundary fields,
satisfying the phantom roots criterion. Namely, we have two invariant
subspaces, $G_M^{+}$ and $G_M^{-}$, and the eigenvectors of $H$ for each
subspace will be given by separate CCBA.  Furthermore, the Bethe eigenvectors
will be given not as a linear combination of independent original basis
vectors, but as a linear combination of the original basis vectors plus other
extra auxiliary vectors, which are linearly dependent and are added for
convenience.  Adding the extra vectors allows to symmetrize the basis and to make
the CCBA coefficients elegant and simple. Below we remind of the definition of
the basis vectors and show how the extra auxiliary vectors are constructed.

Define the following local left vectors on each site $n$ 
\bee 
&&\phi_n(x)=\left(1,\,-\eE^{\th_-+\al_-+\be_-+(2x-n+1)\eta}\right)\equiv\left(1,\,\eE^{z_{n,x}}\right),\label{left;local}\\
&&z_{n,x}=\th_-+\al_-+\be_-+(2x-n+1)\eta+\ir\pi.\label{Phase}
\eee 
Here the second component of these states depends on the position index $n$
and $z_{n,x}$ serves as a phase factor of the state $\phi_n(x)$. Let us
introduce a set of factorized states

\begin{align}
	&\bra{\,\underbrace{0,\dots,0}_{m_0},\underbrace{n_1,\dots,n_{k}}_{k}   ,\underbrace{N,\dots,N}_{m_N}}\no\\
	&=\eE^{\eta (N m_N+\sum_{j=1}^k n_j)}\bigotimes_{l_1=1}^{n_1}\phi_{l_1}(m_0)\bigotimes_{l_2=n_1+1}^{n_2}\phi_{l_2}(m_0+1)\no
\\
	&\quad\dots
	\bigotimes_{l_{k+1}=n_k+1}^{N}\phi_{l_{k+1}}(m_0+k),\label{BraVecs}\\
	&0<n_1<n_2<\cdots<n_k<N, \quad k\geq 0.\no
	\end{align}

\begin{figure}[htbp]
	\centerline{
		\includegraphics[width=0.45\textwidth]{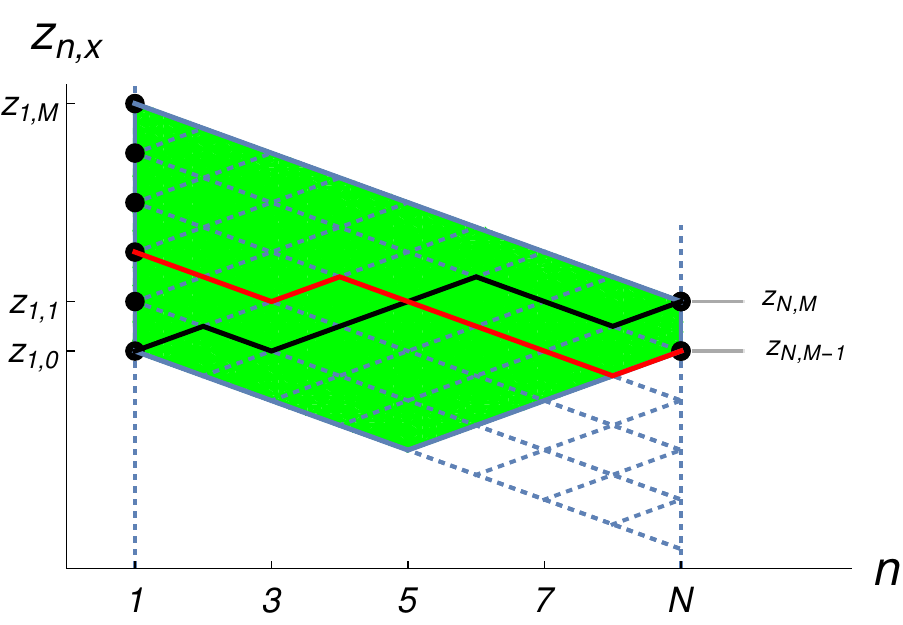}
	}
	\centerline{
		\includegraphics[width=0.45\textwidth]{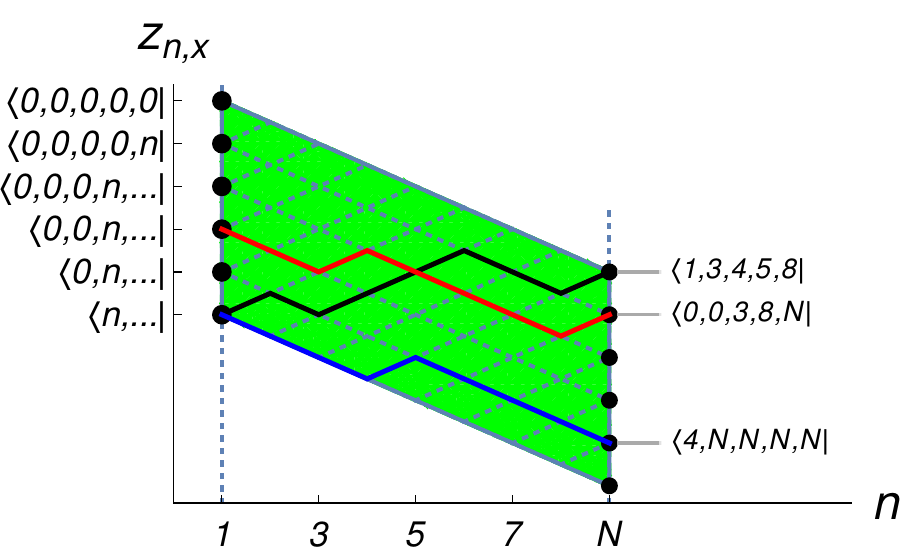}
	}
	\caption{Visualization of the invariant subspace $G_M^{+}$ (Upper
          Panel) and of the symmetrically enlarged $G_M^{+}$ with auxiliary
          states added (Lower Panel) for $N=9, M=5$, and showing the phase
          factor $z_{n,x}$ from (\ref{left;local}) versus site number $n$. Any
          state (\ref{BraVecs}) corresponds to some directed path (backward
          moves are forbidden).  \textbf{Upper Panel}: Illustration of
            the linearly independent states (\ref{IndependentBasis}) that
            realize a basis of $G_M^{+}$.  Directed paths start at one of
          $M+1$ points (filled black circles) on site $n=1$ and end at one of
          two points indicated by filled circles at $n=N$.  The allowed
            paths representing the basis states (\ref{IndependentBasis}) lie
            entirely inside the filled region, including the boundaries. The
            black and red trajectories are examples of two states from
            (\ref{IndependentBasis}): $\bra{1,3,4,5,8}$ and
          $\bra{0,0,3,8,N}$, respectively.  \textbf{Lower Panel}: Illustration
          of the full set of states entering the chiral coordinate Bethe
          ansatz (\ref{CBA}). Directed paths start at one of $M+1$ filled
          circles on site $n=1$ and end at one of $M+1$ filled circles at
          $n=N$. The blue line represents a state $\bra{4,N,N,N,N}$ which
          belongs to the extra set of auxiliary states (\ref{Extra;States}),
          while the black and the red line ``belong" to the original set of
          basis states, see Upper Panel.  }
	\label{Fig-Mgrid}
\end{figure}

The structure of the states (\ref{BraVecs}) is particular and is very
  different from the usual computational basis of up and down spins, used for
instance to describe the Bethe eigenstates of a periodic XXZ spin chain.  The
number $m_0$ defines the initial phase of the first qubit, and the phases of
the subsequent qubits increment by an amount $\eta$ from site to site except
at the points $n_1, \ldots n_k$, where kinks occur. The states (\ref{BraVecs})
are conveniently graphically represented in a form of trajectories, see
Fig.~\ref{Fig-Mgrid}.  The nature of any state even in the presence of
  kinks is chiral.  The full set of Bethe vectors (all eigenstates of the
Hamiltonian) will be expressed by a chiral set (\ref{BraVecs}) as explained
below.

It was proved in \cite{PhantomLong} that the  bra vectors (\ref{BraVecs}) with
\begin{align}
& m_0+k+ m_N = M, \no \\
&m_0=0,1,\ldots M, \quad  m_N=0,1,  \label{IndependentBasis}
\end{align}
are all independent and form a basis of the invariant subspace $G_M^{+}$, with
the dimension $\dim G_M^{+} = d_{+}(M)=\sum_{n=0}^M\binom{N}{n}$.  The
Hamiltonian $H$ has $d_{+}(M)$ left eigenvectors which are linear combinations
of the $G_M^{+}$ basis states.  The $G_M^{+}$ basis (\ref{IndependentBasis})
consists of factorized states with 0 kink, 1 kink, etc. $\dots$ up to $M$
kinks, see Fig.~\ref{Fig-Mgrid}, Upper Panel.


For our purpose it is convenient to enlarge the basis by adding to
(\ref{IndependentBasis}) extra chiral states of the form (\ref{BraVecs}) with
\begin{align}
&m_0+k+ m_N = M,\no\\
&m_0=0,1,\ldots M,\quad m_N=2,\ldots M, \label{Extra;States}
\end{align}
rendering the enlarged set of states 
\begin{align}
&m_0,m_N =0,1,\ldots M, \quad m_0+k+ m_N = M, \label{Total;States}
\end{align}
completely symmetric, see Fig.~\ref{Fig-Mgrid}, Lower panel.
For $M=0$ and $M=1$, the basis vector set (\ref{IndependentBasis}) coincides with the enlarged set (\ref{Total;States}). 
For $M>1$ the number of auxiliary vectors increases monotonically with $M$.
For $M=2,3,4$, the number of auxiliary vectors is $1, N+1, \frac{N^2+N+4}{2}$ respectively. For arbitrary $M\geq 2$,
the number of additional vectors can be calculated on combinatorial grounds and is equal to 
\begin{align} 
d_+^{add}(M)&=\sum_{j=0}^{M-2}\sum_{k=0}^{j}\binom{N-1}{k}.
\end{align} 
It can be proved (see Appendix B) that all auxiliary vectors are linear
combinations of the $d_{+}(M)$ basis vectors.  The full generating set
(\ref{Total;States}) contains in total
\begin{align}
d_+^{total}(M)&=\sum_{j=0}^M\sum_{k=0}^{j}\binom{N-1}{k}\label{DplusTotal}
\end{align}
vectors. Each vector from the set corresponds to a directed path in
Fig.~\ref{Fig-Mgrid}, Lower Panel.


Note that in the $G_M^+$ case, we deal with the bra vectors. In the following
we show how to construct the auxiliary vectors for the ket $G_M^-$ basis.

\subsection{Adding auxiliary ket vectors to the basis of $G_M^-$}
Analogously, introduce the  local ket states,
\bee
\tilde\phi_n(x)=\left(\begin{array}{c}
	1\\
	\eE^{-\th_--\al_--\be_-+(2x-n+1)\eta}
\end{array}
\right),\label{right;local}
\eee
and construct factorized  states out of them 
\begin{align}
&\ket{\,\underbrace{0,\dots,0}_{m_0},\underbrace{n_1,\dots,n_{k}}_{k} ,\underbrace{N,\dots,N}_{m_N}}\!\rangle,\no
\end{align}
obtainable from bra vectors (\ref{BraVecs}) via the replacement $\phi
\rightarrow \tilde{\phi}$.  Analogously to (\ref{IndependentBasis}), the above
ket states with
\begin{align}
& m_0+k+ m_N = \tM,\no \\
&m_0=0,1,\ldots \tM, \quad  m_N=0,1,  \label{M;RightBasis}
\end{align}
where $\tM =N-1-M$, form a basis of the invariant subspace $G_M^-$
\cite{PhantomLong}. Adding additional ket states in analogy to
(\ref{Extra;States}), we get another fully symmetric set of ket vectors with
\begin{align}
&m_0,m_N =0,1,\ldots \tM,\quad m_0+k+ m_N = \tM,\label{M;RightBasisTotal}
\end{align}
and their total number is 
\bee
d_-^{total}(M)=\sum_{j=0}^{\tM}\sum_{k=0}^{j}\binom{N-1}{k}.
\eee

\section{Phantom Bethe Eigenstates in $G_M^{+}$ for  $M=0,1,2$}

\subsection{$M=0$ case}

When $M=0$, the invariant subspace $G_0^+$ consists of just one state,  a spin-helix state (SHS) \cite{2016PopkovPresilla,2017PopkovSchutzHelix,Pozsgay2021}
\bee
\langle\Psi_0|=\phi_1(0)\cdots\phi_N(0), \label{SHS}
\eee with 
\bee
&&\bra{\Psi_0}H=\bra{\Psi_0}E_0,\quad \\
&&E_0=-\sh\eta(\coth(\al_-)+\tanh(\be_-)\no\\
&&\qquad\,\,+\coth(\al_+)+\tanh(\be_+)),\label{E0}
\eee
see \cite{PhantomShort,PhantomLong}.  In the factorized state $\bra{\Psi_0}$, 
the qubit phase grows linearly, implying underlying chiral properties of the state.
Indeed, for a Hermitian Hamiltonian in the easy plane regime, the SHS $\bra{\Psi_0}$ 
for $\be_{\pm}=0$ carries the magnetic current
\begin{align}
&j^z=\frac{\bra{\Psi_0}{\mathbf j}_k^z\ket{\Psi_0}}{\langle\Psi_0|\Psi_0\rangle}=2\sin\gamma, \label{Current;SHS}\\
&{\mathbf j}_k^z=2(\sigma^x_k\sigma^y_{k+1}-\sigma^y_k\sigma^x_{k+1}).\no
\end{align}
For a Hermitian system in the easy axis regime (\ref{hermiticity_2}), the SHS
$\bra{\Psi_0}$ carries no magnetic current, i.e.~$j^z=0$.  Remarkably, the
  SHS (\ref{SHS}) has been produced experimentally in a system of cold atoms
  where the $z$-axis anisotropy of the Heisenberg interaction can be controlled by
  Feshbach resonance \cite{2020NatureSpinHelix,Jepsen2021}.  

\subsection{$M=1$ case}

Define the following factorized states 
\bee
\langle n|=\eE^{n\eta}\phi_1(0)\cdots\phi_{n}(0)\phi_{n+1}(1)\cdots
\phi_N(1).\label{basis_1}
\eee
The states $\langle 0|,\,\langle 1|,\dots,\langle N|$ span the subspace $G_1^+$ \cite{PhantomLong}. Consequently there 
exist $N+1$ Bethe eigenstates which are linear combinations of the basis vectors
\bee
\langle\Psi_{1}^{(\al)}|=\sum_{n=0}^N\langle n|f_{n}^{(\al)}, \quad \al = 0,1,\ldots, N,
\eee
where the greek upper index $\al$ enumerates the states of the $G_1^{+}$ multiplet.

Define the boundary parameters
\bee 
a_{\pm}=\frac{\sh(\al_\pm+\eta)}{\sh(\al_{\pm})},\quad b_{\pm}=\frac{\ch(\be_\pm+\eta)}{\ch(\be_{\pm})}.\label{def;ab}
\eee
The coefficients $\{f_n^{(\al)}\}$ can be written in the following coordinate
Bethe ansatz form \cite{PhantomLong}
\bee 
&&f_{n}^{(\al)}=g_n\left(A_+^{(\al)}\eE^{\ir n\rP(\al)}+A_-^{(\al)}\eE^{-\ir n\rP(\al)} \right),\quad 0\leq n\leq N,\no\\
&&g_0=\frac{1}{1-a_-b_-},\quad g_N=\frac{1}{1-a_+b_+},\no\\
&&g_1=g_2=\dots=g_{N-1}=1.
\label{M1;ansatz}
\eee
Note that writing $f_n^{(\al)}$ as a product of the listed $g_n$ times a
  second factor allows this one to be a sum of plane waves for all sites $n$
  even at the ends with $n=0$ and $N$.
The quasi-momentum $\rP(\al)$ is subject to Eq.~(\ref{BAE-M1}) which is
the consistency condition for the following relations for {the amplitudes $A^{(\alpha)}_{\pm}$
\bee 
&&A^{(\alpha)}_{-}=S_L(\rP(\alpha))A^{(\alpha)}_{+},\no
\\
&&A^{(\alpha)}_{-}=\eE^{2\ir N \rP(\alpha)}S_R(\rP(\alpha))A^{(\alpha)}_{+},\label{M1;A-+}
\eee
where $S_L(\rP)$ and $S_R(\rP)$ are the reflection matrices on left and right
boundaries
\cite{Note1} respectively with
\bee 
&&S_L(\rP)=-\frac{1-a_-\,\eE^{\ir\rP}}{a_--\eE^{\ir\rP}}\,\,\frac{1-b_-\,\eE^{\ir\rP}}{b_--\eE^{\ir\rP}},\label{RM_1}\\
&&S_R(\rP)=-\frac{a_+-\eE^{\ir\rP}}{1-a_+\,\eE^{\ir\rP}}\,\,\frac{b_+-\eE^{\ir\rP}}{1-b_+\,\eE^{\ir\rP}}.\label{RM_2}
\eee
}
The compatibility condition of Eq.~(\ref{M1;A-+}) is exactly the BAE for $M=1$
\begin{align}
\eE^{2\ir  N\rP}\,\prod_{\sigma=\pm}\frac{a_\sigma-\eE^{\ir\rP}}{1-a_\sigma\,\eE^{\ir\rP}}\,\frac{b_\sigma-\eE^{\ir\rP}}{1-b_\sigma\,\eE^{\ir\rP}}=1. \label{BAE-M1}
\end{align}
The solutions of BAE (\ref{BAE-M1}) are denoted by $\rP(\al)$ with $\al=0,\dots,N$.
The corresponding eigenvalue in terms of the  Bethe root $\rP(\al)$ is given by
\begin{align}
E(\al)=4\cos(\rP(\al))-4\De+E_0.
\end{align}
For a Hermitian system, the single quasi-momentum $\rP(\al)$ can be real or purely imaginary.
It has been proved in \cite{PhantomLong} that the invariant subspaces $G_1^+$ have additional internal structure when at least one of the additional constraints $a_\pm b_\pm=1$ is satisfied. 

Once the eigenstates are constructed,  physical quantities can be
calculated, e.g.~the expectation value of the spin current. A qualitative
  analysis yields that the spin currents in the single particle multiplet can
differ from the SHS current $j_{SHS}^z=2\sin\gamma$ at most by
$O\left(\frac1N\right)$ corrections in the easy plane regime.  Consider a
Hermitian Hamiltonian in the easy plane regime with the  boundary
parameters
\begin{align}
&\be_+=\be_-=0,\quad \al_\pm=-\ir\gamma\pm\ir\frac{\pi}{2}\quad \mbox{mod}\,\, 2\pi\ir,\no\\
&\th_{-} - \th_{+} = \ir(N-1)\gamma \mod 2\pi \ir.
\end{align}
The explicit expressions of the current in the $N+1$ eigenstates are \cite{PhantomLong} 
\begin{align}
{j}^z(\al)&=\frac{\langle \Psi_1^{(\al)}|\,\mathbf{j}_l^z|\Psi_1^{(\al)}\rangle}{\langle \Psi_1^{(\al)}| \Psi_1^{(\al)}\rangle}\no\\
&=2\sin\gamma\left(1-\frac{4}{N}\frac{1-\cos^2 (\rP(\al))}{1+\Delta^2-2\Delta\cos(\rP(\al))}\right),\label{current_case(iii)}\\
\rP(\al)&=\frac{\pi \al}{N},\quad \al=0,\ldots,N.\no
\end{align}

It can be seen from the above that all phantom Bethe states are current carrying states: 
 the upper and lower bounds for the current of the multiplet are
 of order of the SHS current $ j_{SHS}$,
\begin{align}
  j_{SHS} \left(1-\frac{4}{N}\right)  \leq   j^z(\al)  \leq j_{SHS}=2 \sin \ga.
\end{align}
The upper bound is saturated; indeed $j^z(0) = j^z(N) = j_{SHS}$ since the
respective Bethe states $\langle \Psi_{1}^{(\al)}|$ with $\al=0,N$ are in fact
spin helix states, differing by an initial phase. The lower bound is
approached most closely for $\rP(\al)\leq\gamma\leq \rP(\al+1)$, and it can be
saturated if an $\alpha$ satisfies $\rP(\al)=\ga$, i.e.~for some root of unity
anisotropies.

In the following we omit the upper index $\al$ enumerating the physical BAE
solutions for brevity of notation.

\subsection{$M=2$ case}
For $M=2$, we follow the same procedure to construct the Bethe eigenstates, via a generating set 
(\ref{Total;States}), i.e.~vectors   $\bra{0,0}, \bra{0,1}, \ldots , \bra{0,N}, \ldots , \bra{N,N}$.

Using convenient notations 
\bee
w_{\pm}=a_\pm+b_\pm,\label{def;w}
\eee 
the action of $H$ on the set $\bra{n,m}$ is given by
\begin{widetext}
\bee
&&\langle 0,0|H =\left(4\De\, a_-b_--w_--w_+\right)\langle 0,0|+4\De\,(w_--2\De\,\,a_-b_-)\langle 0,1|,\label{H00}\\[2pt]
&&\langle 0,1|H=\left(w_--w_+-4\De\,a_-b_-\right)\langle 0,1|+2\langle 0,2|+2\,a_-b_-\langle 0,0|,\\[2pt]
&&\langle 0,n|H=\left(w_--w_+-4\De\right)\langle 0,n| +2\langle0,n\!-\!1|+2\langle0,n\!+\!1|+2(1-a_-b_-)\langle 1,n|,\quad 2\leq n\leq N\!-\!1,\\[2pt]
&&\langle 0,N|H=\left(w_-+w_+-4\De\right)\langle 0,N|+2(1-a_-b_-)\langle1,N|+2(1-a_+b_+)\langle0,N\!-\!1|,\label{H0N}\\[2pt]
&&\langle n,m|H=-\left(w_-+w_++4\De\right)\langle n,m|+2\langle n\!-\!1,m|+2\langle n\!+\!1,m|\no\\
&&\hspace{1.8cm}+2\langle n,m\!+\!1|+2\langle n,m\!-\!1|,\quad 1\!\leq\!n\!<\!m\!\leq\!N\!-\!1,\quad m\!-\!n\!>\!1,
\\[2pt]
&&\langle n,n\!+\!1|H=-\left(w_-+w_+\right)\langle n,n\!+\!1|+2\langle n\!-\!1,n\!+\!1|+2\langle n,n\!+\!2|,\quad 1\!\leq\!n\!\leq\!N\!-\!2,\\[2pt]
&&\langle n,N|H=\left(w_+-w_--4\De\right)\langle n,N|+2\langle n\!-\!1,N|+2\langle n\!+\!1,N|+2(1-a_+b_+)\langle n,N\!-\!1|,\,\, 1\!\leq\!n\!\leq\!N\!-\!2,\,\,\,\\[2pt]
&&\langle N\!-\!1,N|H=\left(w_+-w_--4\De \,a_+b_+\right)\langle N\!-\!1,N|+2\langle N\!-\!2,N|+2a_+b_+\langle N,N|,\\[2pt]
&&\langle N,N|H =\left(4\De a_+b_+-w_+-w_-\right)\langle N,N|+4\De(w_+-2\De\,a_+b_+)\langle N\!-\!1,N|.\label{HNN}
\eee
\end{widetext}
Obviously, the factorized states $\bra{n,m}$ span an invariant subspace  $G_2^+$of $H$.  
The respective phantom Bethe eigenstates belonging to $G_2^+$ can be written as a linear combination of   $\bra{n,m}$ as
\begin{align} 
\langle \Psi_2|=&\sum_{0\leq n_1<n_2\leq N}\langle n_1,n_2|\,{f}_{n_1,n_2}+\sum_{n=0,N}\bra{n,n}f_{n,n},
\end{align}
with yet unknown eigenvalue $E$. Later, for convenience, we extend the
notation to a double sum over $0\leq n_1\leq n_2\leq N$ with, however,
$f_{n,n}\equiv 0$ for $n\neq 0,N$.

We write $E$ as
\bee
E=2\Lambda-8\De +E_0,
\eee 
where $E_0$ is defined in Eq.~(\ref{E0}).
The eigenvalue equation $\langle\Psi_2| H =\langle \Psi_2|E$ gives rise to the following recursive identities for the  coefficients $f_{n,m},$
\begin{widetext}
\bee
&&(\Lambda-2\De\delta_{n+1,m})f_{n,m}=f_{n+1,m}+f_{n-1,m}+f_{n,m+1}+f_{n,m-1},\quad 2\leq n<m\leq N-2,\label{f;nm}\\
&&\left(\Lambda-2\De\delta_{n, N-2}\right)f_{n,N-1}=f_{n+1,N-1}\!+\!f_{n-1,N-1}\!+\!f_{n,N- 2}\!+\!(1\!-\!a_+b_+)f_{n,N},\quad 2\!\leq\! n\!\leq\!  N\!-\!2,\label{f;nN-1}\\[2pt]
&&(\Lambda-2\De\delta_{2,m})f_{1,m}=f_{1,m+1}+f_{1,m-1}+f_{2,m}+(1-a_-b_-)f_{0,m},\quad 2\leq m\leq N-2,\label{f;1m}\\ [2pt]
&&\Lambda\, f_{1,N-1}=f_{2,N-1}+f_{1,N-2}+(1-a_+b_+)f_{1,N}+(1-a_-b_-)f_{0,N- 1},\label{f;1N-1}\\ [2pt]
&&\left(\Lambda-w_-
\right)f_{0,m}=f_{0,m-1}+f_{0,m+1}+f_{1,m},\quad 2\leq m \leq N-2.\qquad\label{f;0m}\\ [2pt]
&&\left(\Lambda-w_+
\right)f_{n,N}=f_{n-1,N}+f_{n+1,N}+f_{n,N- 1},\quad 2\leq n \leq N-2,\label{f;nN},\\ [2pt]
&&\left(\Lambda-w_-\right)f_{0,N- 1}=(1-a_+b_+)f_{0,N}+f_{0,N- 2}+f_{1,N- 1},\label{f;0N-1}\\[2pt] 
&&\left(\Lambda-w_+
\right)f_{1,N}=(1-a_-b_-)f_{0,N}+f_{2,N}+f_{1,N- 1},\label{f;1N}\\ [2pt]
&&\left(\Lambda-w_--w_+\right)f_{0,N}=f_{0,N- 1}+f_{1,N}.\label{f;0N}\\ [2pt]
&&(\Lambda+2\De\,a_-b_--w_--2\De)f_{0,1}=f_{0,2}+2\De(w_--2\De\,a_-b_-)f_{0,0},\label{f;01}\\ [2pt]
&&(\Lambda+2\De\,a_+b_+-w_+-2\De)f_{N- 1,N}=f_{N- 2,N}+2\De(w_+-2\De\,a_+b_+)f_{N,N },\label{f;N-1N}\\ [2pt]
&&\left(\Lambda-2\De\, a_-b_--2\De\right)f_{0,0}=a_-b_-f_{0,1},\label{f;00}\\[2pt]
&&\left(\Lambda-2\De\, a_+b_+-{2\De}\right)f_{N,N}=a_+b_+f_{N- 1,N}.\label{f;NN}
\eee
\end{widetext}
We propose the following ansatz 
{\begin{align}
f_{n,m}=g_{n,m}\sum_{\sigma_1,\sigma_2=\pm}\left( A^{1,2}_{\sigma_1,\sigma_2}\,\eE^{\ir \sigma_1n \rP_1+\ir \sigma_2m\rP_2}\right.\no\\
\left. +{A}^{2,1}_{\sigma_2,\sigma_1}\,\eE^{\ir \sigma_2n \rP_2+\ir \sigma_1m\rP_1}\right),\label{M2;ansatz}
\end{align}}
where $\rP_1,\rP_2$ are quasi-momenta and the coefficients $\{g_{n,m}\}$ are
$\rP$-independent.  We impose $g_{n,m}\equiv1$ for $n,m\neq 0,N$.
Considering the bulk term Eq.~(\ref{f;nm}) with $m\neq n+1$ and using the
ansatz (\ref{M2;ansatz}), we get the expression of $\Lambda$ and energy
\begin{align}
&\Lambda=2\cos(\rP_1)+2\cos(\rP_2),\label{Lambda}\\
&E=4\sum_{j=1}^2\cos(\rP_j)-8\De+E_0.
\end{align} 
To satisfy Eq.~(\ref{f;nm}) with $m=n+1$, we get the two-body scattering
matrix \cite{Note1}
{\begin{align}
A^{2,1}_{\sigma_2,\sigma_1}=S_{1,2}(\sigma_1\rP_1,\sigma_2\rP_2)A^{1,2}_{\sigma_1,\sigma_2},
\end{align}
where $S$ has the following symmetry and explicit expression
\begin{align}
&S_{1,2}(\rP,\rP')=S_{2,1}(-\rP',-\rP)\no\\
&=-\frac{1-2\De\,\eE^{\ir \rP'}+\eE^{\ir \rP'+\ir \rP}}{1-2\De\,\eE^{\ir \rP}+\eE^{\ir \rP'+\ir\rP}}.\label{S_matrix}
\end{align}}
The ansatz (\ref{M2;ansatz}) allows us to get the following expressions from Eqs.~(\ref{f;nN-1}) and (\ref{f;1m})
\bee 
&&g_{n,N}=\frac{1}{1-a_+b_+},\quad g_{0,n}=\frac{1}{1-a_-b_-},\label{g;0m}\\
&&2\leq n\leq N- 2.\no
\eee
The boundary dependent Eqs.~(\ref{f;0m}) and (\ref{f;nN}) determine the following left and right reflection matrices respectively
\bee 
&&A^{j,k}_{-,\sigma_k}=S_L(\rP_j)A^{j,k}_{+,\sigma_k},\\
&&A^{j,k}_{\sigma_j,-}=\eE^{2\ir N\rP_k}S_R(\rP_k)A^{j,k}_{\sigma_j,+},
\eee
where the reflection matrices $S_L(\rP)$ and $S_R(\rP)$ are given by Eqs. (\ref{RM_1})-(\ref{RM_2}).
\bee 
&&S_L(\rP)=-\frac{1-a_-\,\eE^{\ir\rP}}{a_--\eE^{\ir\rP}}\,\,\frac{1-b_-\,\eE^{\ir\rP}}{b_--\eE^{\ir\rP}},\\
&&S_R(\rP)=-\frac{a_+-\eE^{\ir\rP}}{1-a_+\,\eE^{\ir\rP}}\,\,\frac{b_+-\eE^{\ir\rP}}{1-b_+\,\eE^{\ir\rP}}.
\eee
The scattering matrix in (\ref{S_matrix}) and reflection matrices  in
(\ref{RM_1}), (\ref{RM_2}) determine all the amplitudes
$A^{j,k}_{\sigma_j,\sigma_k}$. The consistency condition of our ansatz 
gives the BAE 
\bee
&&\eE^{2\ir N\rP_j}S_{j,k}(\rP_j,\rP_k)S_R(\rP_j)S_{k,j}(\rP_k,-\rP_j)\no\\
&&\times S_L(-\rP_j)=1,\quad j,k=1,2, \quad j\neq k.\label{BAE;M2}
\eee
One can verify that our BAE in (\ref{BAE;M2}) is consistent with the one
given by the modified ABA \cite{OffDiagonal03} and the functional $T$-$Q$ relation
\cite{Nepomechie2003,OffDiagonal}, see Appendix A.  Letting $m$ in
(\ref{f;0m}) and $n$ in (\ref{f;nN}) take values 2 and $N-2$ respectively and
using the reflection matrices (\ref{RM_1}), (\ref{RM_2}), we have
\begin{align}
g_{1,N}&=g_{N-1,N}=\frac{1}{1-a_+b_+},\no\\
g_{0,1}&=g_{0,N-1}=\frac{1}{1-a_-b_-},\label{g;01}
\end{align}
 extending the  result (\ref{g;0m}) to $1\leq n\leq N- 1$.
Substituting the result in (\ref{g;01}) into Eq.~(\ref{f;0N-1}), we get the expression of $g_{0,N}$
\bee 
g_{0,N}=\frac{1}{(1\!-\!a_-b_-)(1\!-\!a_+b_+)}.\label{g;0N}
\eee
The remaining coefficients $f_{0,0}$ and $f_{N,N}$ are derived from Eqs.~(\ref{f;01}) and (\ref{f;N-1N})
\begin{align}
g_{0,0}&=\frac{a_-b_-}{2\Delta(1\!-\!a_-b_-)(w_--\!2\De\,a_-b_-)},
\no\\ g_{N,N}&=\frac{a_+b_+}{2\Delta(1\!-\!a_+b_+)(w_+\!-\!2\De\,a_+b_+)}.\label{gNN}
\end{align}

Using Eqs. (\ref{Parameterization;1})-(\ref{Parameterization;3}), we reparameterize the  functions $\{g_{n,m}\}$ in terms of $\al_\pm$ and $\be_\pm$ as
\bee
g_{n,m}=\begin{cases}
	1, & n,m\neq 0,N,\\[2pt]
	F_-(1), & n=0,\,\,\, m\neq 0,N,\\[2pt]
	F_+(1), & n\neq 0, N,\,\,\, m=N,\\[2pt]
	F_-(1)F_+(1), & n=0,\,\,\,m=N,\\[2pt]
	F_-(1)F_-(2), & n=m=0, \\[2pt]
	F_+(1)F_+(2), & n=m=N,
\end{cases}\label{g}
\eee
where
\begin{align}
F_{\si}(k) =&\,\, \delta_{k,0}-(1-\delta_{k,0})\frac{\sh(\al_\si+(k-1)\eta)}{\sh(k\eta)},\no\\
&\,\,\times \frac{\ch(\be_\si+(k-1)\eta)}{\ch(\al_\si+\be_\si+k\eta)},\quad \sigma=\pm.
\label{F;function}
\end{align}
Note that for our futher generalization (\ref{gCoeff}) it is convenient 
to define  $F_\si(0)=1$ via (\ref{F;function}),
even though $F_\si(0)$ does not appear in (\ref{g}).
One can prove that our ansatz (\ref{M2;ansatz}), (\ref{g}) satisfies all the
relations (\ref{f;nm})-(\ref{f;NN}), see Appendix D.

\textit{Remark. }
In the generic case, the invariant subspace $G_2^+$ is irreducible. However, on special
manifolds, further internal structures appear, leading to the existence of one or
more sub-subspaces, which are invariant w.r.t.~the action of the Hamiltonian. 
As an example, for  $a_\pm b_\pm=1$, three invariant subspaces of $G_M^{+}$  appear.
The details of this  further structuring  and the consequences for the
BAE sets is discussed in Appendix E.

\bigskip

\section{Generalization for arbitrary $M$}
On the basis of our findings we formulate the following hypothesis: phantom
Bethe vectors, i.e.~Bethe states with infinite rapidities resp.~momenta
$k_j=\pm\gamma$, are for general $M$ given by a superposition of the
  states (\ref{BraVecs}). We denote the vector $\bra{\,0,\dots,0,\tilde
    n_1,\dots,\tilde n_{k}, N,\dots,N}$ from (\ref{BraVecs}) simply as
  $\bra{n_1, \dots, n_M}$, where some of the first site labels $n_j$ may be
  identical to 0 and some of the last ones identical to $N$.

We have seen in the $M=1, 2$ cases, that writing $f_n$ or $f_{n,m}$ as a
product of certain prefactors $g_n$ or $g_{n,m}$ times a second factor allows
this one to be a sum of plane waves for all sites even at the ends with $n$ or
$m$ equal to $0$ or $N$.  Analysing the $M=1, 2$ cases, we see that the
prefactors $g$ only depend on the number of site labels 0 resp.~$N$.  This
inspired us to formulate a general rule for the arbitrary $M$ case with a
certain prefactor $g_{n_1, \dots , n_M}\equiv C_{m_0,m_N}$ where $m_0$ and
$m_N$ denote the number of site labels equal to 0 resp.~$N$ in the sequence
$n_1, \dots , n_M$.  Using this rule we find
\begin{align}
&\bra{\Psi_M} =\sum_{n_1, \dots , n_M}  \bra{n_1, \dots, n_M} f_{n_1,\dots,n_M}, \label{CBA}\\
&f_{n_1,\ldots,n_M}=C_{m_0,m_N}\sum_{r_1,\ldots,r_M}\,\sum_{\sigma_1,\ldots,\sigma_M=\pm}A^{r_1,\ldots,r_M}_{\sigma_{r_1},\ldots,\sigma_{r_M}}\no\\
&\hspace{1.8cm}\times\eE^{\ir \sum_{k=1}^M\sigma_{r_k}n_k\rP_{r_k}},\label{M;ansatz}
\end{align}
where in (\ref{CBA}) we sum over all configurations $n_1, \dots , n_M$ allowed
by (\ref{Total;States}).  The first sum in (\ref{M;ansatz}) is over all
permutations $r_1,\dots,r_M$ of $1,\dots,M$, while the coefficients
$C_{m_0,m_N}$ depend only on $m_0,m_N$ and are given by remarkably simple
expressions
\begin{align}
C_{m_0,m_N}=\prod_{k=0}^{m_0} F_{-}(k) \prod_{l=0}^{m_N} F_{+}(l). \label{gCoeff}
\end{align}
where $F_\sigma(m)$  are defined by Eq.~(\ref{F;function}).
The amplitudes $A^{r_1,\dots r_M}_{\sigma_{r_1},\dots,\sigma_{r_M}}$ 
are determined by the two-body scattering matrix $S$ in (\ref{S_matrix})
  and the reflection matrices $S_L$, $S_R$ in (\ref{RM_1})-(\ref{RM_2})
\begin{align}
A^{\ldots,r_{n+1},r_{n},\ldots}_{\ldots,\sigma_{r_{n+1}},\sigma_{r_{n}}\ldots}=&\,\,S_{r_n,r_{n+1}}(\sigma_n\rP_n,\sigma_{n+1}\rP_{n+1})\no\\
&\,\,\times A^{\ldots,r_{n},r_{n+1},\ldots}_{\ldots,\sigma_{r_{n}},\sigma_{r_{n+1}}\ldots},\label{M;S-matrix}\\[2pt]
&\hspace{-2.6cm}{A^{r_1,\ldots}_{-,\dots}}=S_L(\rP_{r_1}){A^{r_1,\dots}_{+,\dots}},\label{M;RM_1}\\[2pt]
&\hspace{-2.6cm}A^{\dots,r_M}_{\dots,-}=\eE^{2N\ir \rP_{r_M}}S_R(\rP_{r_M})A^{\dots,r_M}_{\dots,+}.\label{M;RM_2}
\end{align}
The compatibility of the whole scheme is guaranteed by a set of transcendental equations for the quasi-momenta, the BAE
\begin{align}
&\,\,\eE^{2\ir N\rP_{r_1}}S_{r_1,r_2}(\rP_{r_1},\rP_{r_2})\cdots S_{r_1,r_M}(\rP_{r_1},\rP_{r_M})S_{R}(\rP_{r_1})\no\\
&\times S_{r_M,r_1}(\rP_{r_M},-\rP_{r_1})\cdots S_{r_2,r_1}(\rP_{r_2},-\rP_{r_1})S_L(-\rP_{r_1})=1, \no\\
&\,\, r_1=1,\ldots,M.\label{M;BAE}
\end{align}
The BAE (\ref{M;BAE}) coincide with those obtained by other approaches
\cite{OffDiagonal03,Nepomechie2003}. The corresponding eigenvalue in terms of
quasimomenta $\{\rP_1,\dots,\rP_M\}$ is
\begin{align}
E=4\sum_{j=1}^M(\cos(\rP_j)-\De)+E_0.\label{M;Energy}
\end{align}

Analogously we construct the other set of eigenstates $\ket{\Psi_M}\!\rangle$ belonging to $G_M^-$.
The substitutions \begin{align}
\al_{\pm}\rightarrow -\al_{\pm},\quad 
\be_{\pm}\rightarrow -\be_{\pm}, \quad
\th_{\pm}\rightarrow  \ir \pi+ \th_{\pm},\label{Substitution;1}
\end{align}
leave the Hamiltonian invariant and give the following replacements
\bee 
M\to\tM,\quad a_\pm\to \tilde a_\pm,\quad b_\pm\to\tilde b_\pm,\label{Substitution;2}
\eee
where 
\bee 
\tilde a_{\pm}=\frac{\sh(\al_\pm-\eta)}{\sh(\al_{\pm})},\quad \tilde b_{\pm}=\frac{\ch(\be_\pm-\eta)}{\ch(\be_{\pm})}.\label{def;ab2}
\eee
The vectors in (\ref{Total;States}), (\ref{M;RightBasisTotal}), and the
fundamental relations (\ref{bulk_1})-(\ref{psi;phi2}) all show the symmetry
(\ref{Substitution;1}) and (\ref{Substitution;2}).
This is sufficient to prove that the eigenstates $\ket{\Psi_M}\!\rangle$
can be constructed in analogy to (\ref{CBA}).
Following (\ref{CBA}), we make the ansatz
\begin{align}
&\ket{\Psi_M}\!\rangle=\sum_{n_1, \dots , n_{\tM}} \tilde f_{n_1,\dots,n_{\tM}} \ket{n_1, \dots, n_\tM}\!\rangle,\label{Right;CBA}
\end{align}
with
\begin{align}
 \tilde f_{n_1,\ldots,n_\tM}=&\,\,\widetilde C_{m_0,m_N}\sum_{r_1,\ldots,r_\tM}\sum_{\sigma_1,\ldots,\sigma_\tM=\pm}\widetilde A^{r_1,\ldots,r_\tM}_{\sigma_{r_1},\ldots,\sigma_{r_\tM}}\no\\
&\,\,\times \eE^{\ir \sum_{k=1}^\tM \sigma_{r_k}n_k\tilde{\rP}_{r_k}}.\label{M;ansatz;Right}
\end{align}
Substituting $A^{\cdots}_{\cdots}$, $C_{m_0,m_N}$ and $\{\rP_1,\ldots,\rP_M\}$
in Eqs. (\ref{gCoeff})-(\ref{M;Energy}) with $\widetilde A^{\cdots}_{\cdots}$,
$ \widetilde C_{m_0,m_N}$ and $\{\tilde \rP_1,\ldots,\tilde\rP_{\tM}\}$
respectively and then using the substitutions (\ref{Substitution;1}),
(\ref{Substitution;2}), we get another chiral coordinate Bethe ansatz, now for
the $G_M^-$ Bethe eigenvectors.

The chiral coordinate Bethe ansatz in Eqs.~(\ref{CBA})-(\ref{M;BAE}) and
(\ref{Right;CBA})-(\ref{M;ansatz;Right}) are the main result of this
paper. Eqs.~(\ref{CBA})-(\ref{M;BAE}) give the full set of Bethe vectors for
the $G_M^{+}$ invariant subspace and the dual
Eqs.~(\ref{Right;CBA})-(\ref{M;ansatz;Right}) give the full set of Bethe
vectors for the $G_M^{-}$ invariant subspace, in total, all $2^{N}$ phantom
Bethe vectors.

At present, it is difficult to prove our hypotheses in (\ref{M;ansatz}) and
(\ref{M;ansatz;Right}) completely. However, there are many arguments that
corroborate our hypotheses. On one hand, we retrieve the same BAE which
have been obtained by other approaches. On the other hand, the correctness of
our conjecture for at least a part of the coefficients $f_{n_1,\dots,n_M}$ in
(\ref{M;ansatz}) can be proved for arbitrary $M$.

\section{Spin helix eigenstates}

Among the vectors constituting the $G_M^{+}$ basis plus the auxiliary vectors,
there are $M+1$ linearly independent spin helix states (SHS) of the form
\begin{align}
&\langle {SHS};m|=\bigotimes_{n=1}^N\phi_n(m)\propto \langle
  \underbrace{0,\dots,0}_{m},{\underbrace{N,\dots,N}_{M-m}}|,\\ &\qquad
  m=0,\dots,M,\no
\end{align}
which have the same chirality but different initial qubit phase.

Below we look for
conditions under which these SHS become eigenstates of the Hamiltonian.
Acting by the Hamiltonian $H$ on these SHS and using
Eqs.~(\ref{bulk_1})-(\ref{right}), we find
\begin{widetext}
\begin{align}
\langle {SHS};m|H=&-\left(\frac{\sh\eta\,\ch(\al_-\!+\!\be_-\!+\!2m\eta)}{\sh(\al_-)\ch(\be_-)}+\frac{\sh\eta\,\ch(\al_+\!+\!\be_+\!+\!2(M\!-\!m)\eta)}{\sh(\al_+)\ch(\be_+)}\right)\langle SHS;m|\no\\
&+\frac{2\sh\eta\,\sh((M\!-\!m)\eta)\ch(\al_+\!+\!\be_+\!+\!(M\!-\!m)\eta)}{\sh(\al_+)\ch(\be_+)}\langle SHS;m|\sigma_N^z\no\\
& -\frac{2\sh\eta\,\sh(m\eta)\ch(\al_-\!+\!\be_-\!+\!m\eta)}{\sh(\al_-)\ch(\be_-)}\langle SHS;m|\sigma_1^z.
\label{SHS;H}\end{align}
\end{widetext}

It is clear from the above that   the SHS $\langle {SHS};m|$  becomes  an eigenstate of $H$
 if one or two  additional conditions are satisfied, namely:
\begin{itemize}
	\setlength{\itemsep}{0pt}
	\setlength{\parsep}{0pt}
	\setlength{\parskip}{0pt}
	\item[(i)]  when $\ch(\al_-+\be_-+M\eta)=0$, $\langle {SHS};M|$ is an eigenstate of $H$,
	\item[(ii)]  when $\ch(\al_++\be_++M\eta)=0$, $\langle {SHS};0|$ is an eigenstate of $H$,
	\item[(iii)]  when $\ch(\al_+\!+\!\be_+\!+\!(M\!-\!m)\eta)=0$, $\ch(\al_-\!+\!\be_-\!+\!m\eta)=0$,  $m\neq 0,M$, $\langle {SHS};m|$ {is an eigenstate of $H$},\label{SHS;M}
\end{itemize}
and the corresponding eigenvalues are given by Eq.~(\ref{SHS;H}).



\section{Discussion}

We have analyzed the integrable open XXZ spin-$\frac12$ chain 
satisfying the phantom Bethe roots existence Criterion (PRC), 
\bee
\eta=\frac{\al_-\!+\!\be_-\!+\!\al_+\!+\!\be_+\!+\!\th_-\!-\!\th_+\!+\!2\ir m\pi}{N-2M-1},\label{Special;eta}
\eee
where $m$ is an arbitrary integer, and the integer $M$ has the range $0\leq M
\leq N-1$.  For a Hamiltonian under the PRC (\ref{ConditionPhantomPlus}), the
crossing parameter $\eta$ can only take $N-2M-1$ discrete values with relative
positions equidistant in the complex plane \cite{OffDiagonal}.  Under this
condition, the Hilbert space splits into two invariant subspaces
\cite{PhantomLong} and remarkable singular peaks in the magnetization current
of the associated dissipative quantum system occur \cite{2020ZenoPRL}, which
can now be related to the existence of spin helix eigenstates and their
generalizations in the spectrum of the effective Hamiltonian.

Under the PRC, two conventional BAE with $M$ and $\tM=N-M-1$ regular Bethe
roots appear, which correspond to two invariant subspaces $G_M^+$ and $G_M^-$,
with the dimensions $\mathrm{dim}\, G_M^+=\sum_{k=0}^M\binom{N}{k}$ and
$\mathrm{dim}\, G_M^-=\sum_{k=M+1}^N\binom{N}{k}=2^N -\mathrm{dim}\, G_M^+ $.
     
Our proposed chiral coordinate Bethe ansatz allows to construct the full set
of Bethe eigenstates, separately for $G_M^+$ and $G_M^{-}$, as a linear
combination over a symmetric set of vectors, spanning the respective chiral
invariant subspace.  The set of vectors contains spin helix states with
``kinks". Unlike in the periodic case, we have to treat the non-diagonal
boundary fields which break the magnetization conservation, i.e.~the
$U(1)$ symmetry. The integer $M$ determines the maximum number of ``kinks".
An exciting result is that the expansion coefficients for the open spin chain,
in the chiral basis of SHS with kinks, have a very simple analytic form.
 
We demonstrated that for small $M$, the Bethe eigenstates have some unusual
chiral properties such as high magnetization currents.

Our method can be generalized to other integrable open systems, not
necessarily of quantum origin, such as the asymmetric simple exclusion process
(ASEP) with open boundaries \cite{De2005,Zhang2019}, the spin-1
Fateev-Zamolodchikov model \cite{ZF80} and spin-$s$ integrable systems
\cite{ZF-BAE}. Potentially, a generalization of our results to the XYZ
spin-$\frac12$ chain \cite{Yang2006} might exist, which is a challenging open problem.

The formulation of the chiral coordinate Bethe ansatz has become possible due to
the existence of phantom Bethe roots, which appear both in open and periodically
closed systems \cite{PhantomShort}. 

Another interesting question is how to obtain the eigenstates of non-Hermitian
systems under PRC. Using our bases and the chiral coordinate Bethe ansatz
method, we can always construct the left or right eigenstates which correspond
to one subspace, whether the system is Hermitian or not. For a Hermitian
system the dual states can be directly obtained. If the system is not
Hermitian, the construction of the dual states is still
challenging. A very intuitive example is the one-species ASEP with open
boundary conditions, which belongs to the $M=0$ case. The left steady state of
the Markov matrix is a simple factorized state, while the right steady state has a
very complicated structure, which however can be calculated exactly by other
approaches, the matrix product approach \cite{Derrida1993} or the recursive
approach \cite{Schuetz1993}.

Our results may lay the basis for further analytic studies and may possibly
serve for a new understanding relevant for experimental applications, e.g.~the
experimental realization of the model and eigenstates by techniques
presented in \cite{2020NatureSpinHelix,Jepsen2021}.

\section{Acknowledgments}
Financial support from the Deutsche Forschungsgemeinschaft through DFG project
KL 645/20-1, is gratefully acknowledged.  X.~Z. thanks the Alexander von
Humboldt Foundation for financial support. V.~P. acknowledges support by
European Research Council (ERC) through the advanced Grant No. 694544 --
OMNES.

\bigskip

\appendix 
\section{Appendix A: BAE resulting from modified ABA}

It has been proved under condition (\ref{ConditionPhantomPlus})
there exists a conventional BAE
\cite{OffDiagonal03,Nepomechie2003,OffDiagonal},
\begin{align}
&\left[\frac{\sh(x_j+\frac{\eta}{2})}{\sh (x_j-\frac{\eta}{2})}\right]^{2N}\prod_{\sigma=\pm}\frac{\sh(x_j-\alpha_\sigma-\frac{\eta}{2})}{\sh(x_j+\al_\sigma+\frac{\eta}{2})}\no\\
&\times\frac{\ch(x_j-\beta_\sigma-\frac{\eta}{2})}{\ch(x_j+\beta_\sigma+\frac{\eta}{2})}=\prod_{k\neq j}^M\frac{\sh(x_j-x_k+\eta)\,}{\sh(x_j-x_k-\eta)\,}\no
\\&\times \frac{\sh(x_j+x_k+\eta)}{\sh(x_j+x_k-\eta)},\qquad j=1,\ldots,M.\label{ABAE}
\end{align}		
The above BAE, in terms of  the single particle quasi-momentum $\rP_j$ 
\bee 
\eE^{\ir{\rP}_j}=\frac{\sh\left(x_j+\frac{\eta}{2}\right)}{\sh\left(x_j-\frac{\eta}{2}\right)}\,.
\eee
take the form \cite{PhantomLong}
\begin{align} 
&\eE^{2\ir N{\rP}_j}\prod_{\sigma=\pm}\frac{a_\sigma\,-\eE^{\ir{\rP}_j}}{1-a_\sigma\,\eE^{\ir{\rP}_j}} \,\,\frac{b_\sigma-\eE^{\ir{\rP}_j}}{1-b_\sigma\,\eE^{\ir{\rP}_j}}\no\\
&=\prod_{\sigma=\pm}\prod_{k\neq j}^M\frac{1-2\De\,\eE^{\ir{\rP}_j}+\eE^{\ir{\rP}_j+\ir \sigma{\rP}_k}}{1-2\De\,\eE^{\ir \sigma{\rP}_k}+\eE^{\ir{\rP}_j+\ir \sigma{\rP}_k}},
\,\, \,j\!=\!1,\ldots,M,\label{BAE;CBA}
\end{align}
where $a_\pm,\,b_\pm$ are defined in (\ref{def;ab}).  Valid physical Bethe
roots $\{\rP_1,\dots,\rP_M\}$ satisfy the selection rules $\eE^{\ir\rP_j}\neq
\eE^{\pm\ir \rP_k}, \,\,\eE^{\ir\rP_j}\neq \pm 1$. We see 
that Eq.~(\ref{BAE;CBA}) is identical to our BAE (\ref{M;BAE}) in the main text.
The invariance of the Hamiltonian $H$ w.r.t.~the substitution
(\ref{Substitution;1})
under condition (\ref{ConditionPhantomPlus}) allows to construct another
set of homogeneous BAE by replacing $\al_\pm$, $\be_\pm$ and $M$ in
(\ref{BAE;CBA}) with $-\al_\pm$, $-\be_\pm$ and $\widetilde M=N-1-M$
respectively, see \cite{PhantomLong}. The second set of BAE thus reads
\begin{align}
&\eE^{2\ir N{\tilde \rP}_j}\prod_{\sigma=\pm}\frac{\tilde a_\sigma\,-\eE^{\ir{\tilde \rP}_j}}{1-\tilde a_\sigma\,\eE^{\ir{\tilde \rP}_j}} \,\,\frac{\tilde b_\sigma-\eE^{\ir{\tilde \rP}_j}}{1-\tilde b_\sigma\,\eE^{\ir{\tilde \rP}_j}}\no\\
&=\prod_{\sigma=\pm}\prod_{k\neq j}^{\widetilde M}\frac{1-2\De\,\eE^{\ir{\tilde \rP}_j}+\eE^{\ir{\tilde \rP}_j+\ir \sigma{\tilde \rP}_k}}{1-2\De\,\eE^{\ir \sigma {\tilde \rP}_k}+\eE^{\ir{\tilde \rP}_j+\ir \sigma{\tilde \rP}_k}},
\quad j=1,\ldots,\tM,\label{BAE;CBA2}
\end{align}
where $\tilde a_\pm,\,\tilde b_\pm$ are defined in Eq.~(\ref{def;ab2}).


\section{Appendix B: Linear dependence of the  auxiliary vectors}\label{Additional_vectors} 
Here we show that all extra auxiliary bra vectors participating in the CCBA, are
linear combinations of the basis vectors of $G_M^{+}$, and similarly, all
extra auxiliary ket vectors are linear combinations of the $G_M^{-}$ basis
vectors.  For the proof, it is enough to demonstrate that any bra vector from
the extended (symmetrized) bra set is orthogonal to any ket vector from the
extended (symmetrized) ket set, i.e.~(\ref{Orthogonality}).

To this end, 
define
the function $y(n, v_n, \tilde v_n)$ as
\bee
&&\phi_n(v_n)\tilde\phi_n(\tilde v_n)=1-\eE^{2y(n,v_n,\tilde v_n)\eta},\no\\
&&y(n,v_n,\tilde v_n)=v_n+\tilde v_n-n+1.\label{y;function}
\eee
When $y(n,v_n,\tilde v_n)=0$, the local vectors $\phi_n(v_n)$ and $\tilde\phi_n(\tilde v_n)$ are orthogonal. Introduce the inner products
\begin{align}
&\langle n_1,\dots,n_M|m_1,\dots,m_{\tM}\rangle\!\rangle\no\\
&=\eE^{\eta\sum_{j=1}^M n_j+\eta\sum_{k=1}^{\tM} m_k}\prod_{n=1}^N\left(1-\eE^{2y(n,v_n,\tilde v_n)\eta}\right),
\end{align} 
where $\langle n_1,\dots,n_M|$ belongs to the extended  $G_M^+$ set of vectors 
and $|m_1,\dots,m_{\tM}\rangle\!\rangle$  belongs to the extended  $G_M^+$ set of vectors. Obviously, 
\begin{align} 
&0\leq v_1\leq v_2\leq \dots\leq v_N\leq M,\no\\
&0\leq \tilde v_1\leq \tilde v_2\leq \dots\leq \tilde v_N\leq \tM,\no\\
&v_{n+1}-v_n=0,1,\quad \tilde v_{n+1}-\tilde v_n=0,1,
\end{align}
and 
\bee
&&y(n+1,v_{n+1},\tilde v_{n+1})-y(n,v_n,\tilde v_n)=0,\pm 1,\no\\
&&y(1,v_1,\tilde v_1)\geq 0,\quad y(N,v_N,\tilde v_N)\leq 0.
\eee
So $y(n,v_n , \tilde v_n)=0$ holds at least for one point $n\,(1\leq n\leq N)$ which implies that any pair of vectors $\langle n_1,\dots,n_M|\in G_M^+$ and $|m_1,\dots,m_{\tM}\rangle\!\rangle\in G_M^-$ are orthogonal,

\begin{align}
&\langle n_1,\dots,n_M|m_1,\dots,m_{\tM}\rangle\!\rangle=0.\label{Orthogonality}
\end{align} 

\section{Appendix C: The proof of Eqs.~(\ref{H00})-(\ref{HNN})}
It is easy to prove the following identities:
\begin{widetext}
\bee
&&\phi_n(x)\phi_{n+1}(x)h_{n,n+1}=\sh\eta\,\phi_n(x)\,\phi_{n+1}(x)\,\sigma_n^z-\sh\eta\,\phi_n(x)\,\phi_{n+1}(x)\,\sigma_{n+1}^z,\label{bulk_1}\\[2pt]
&&\phi_n(x\!-\!1)\phi_{n+1}(x)h_{n,n+1}=\sh\eta\,\phi_n(x\!-\!1)\,\phi_{n+1}(x)\,\sigma_{n+1}^z-\sh\eta\,\phi_n(x\!-\!1)\,\phi_{n+1}(x)\,\sigma_n^z,\label{bulk_2}\\[2pt]
&&\phi_1(x)h_1=\frac{\sh\eta}{\sh(\al_-)\ch(\be_-)}\left(\ch(\al_-)\sh(\be_-)-\sh(\al_-\!+\!\be_-\!+\!2x\eta)\right)\phi_1(x)\,\sigma_1^z\no\\[2pt]
&&\hspace{1.7cm}-\frac{\sh\eta\,\ch(\al_-\!+\!\be_-\!+\!2x\eta)}{\sh(\al_-)\ch(\be_-)}\phi_1(x),\label{left}\\[2pt]
&&\phi_N(x)h_N=\frac{\sh\eta}{\sh(\al_+)\ch(\be_+)}(\sh(\al_+\!+\!\be_+\!+\!2(M\!-\!x)\eta)-\ch(\al_+)\sh(\be_+))\phi_N(x)\,\sigma_N^z\no\\[2pt]
&&\hspace{1.9cm}-\frac{\sh\eta\,\ch(\al_+\!+\!\be_+\!+\!2(M\!-\!x)\eta)}{\sh(\al_+)\ch(\be_+)}\phi_N(x).\label{right}\\[2pt]
&&h_{n,n+1}\tilde \phi_n(x)\tilde \phi_{n+1}(x)=\sh\eta\,\sigma_n^z\,\tilde \phi_n(x)\,\tilde\phi_{n+1}(x)-\sh\eta\,\sigma_{n+1}^z\,\tilde\phi_n(x)\,\tilde\phi_{n+1}(x),\label{Right;bulk_1}\\[2pt]
&&h_{n,n+1}\tilde\phi_n(x\!-\!1)\tilde\phi_{n+1}(x)=\sh\eta\,\sigma_{n+1}^z\,\tilde\phi_n(x\!-\!1)\,\tilde\phi_{n+1}(x)-\sh\eta\,\sigma_n^z\,\tilde\phi_n(x\!-\!1)\,\tilde\phi_{n+1}(x)\,,\label{Right;bulk_2}\\[2pt]
&&h_1\,\tilde\phi_1(x)=\frac{\sh\eta}{\sh(\al_-)\ch(\be_-)}\left(\ch(\al_-)\sh(\be_-)-\sh(\al_-\!+\!\be_-\!-\!2x\eta)\right)\sigma_1^z\,\tilde\phi_1(x)\no\\[2pt]
&&\hspace{1.7cm}+\frac{\sh\eta\,\ch(\al_-\!+\!\be_-\!-\!2x\eta)}{\sh(\al_-)\ch(\be_-)}\tilde\phi_1(x),\label{Right;left}\\[2pt]
&&h_N\,\tilde\phi_N(x)=\frac{\sh\eta}{\sh(\al_+)\ch(\be_+)}(\sh(\al_+\!+\!\be_+\!-\!2(\tM\!-\!x)\eta)-\ch(\al_+)\sh(\be_+))\sigma_N^z\,\tilde\phi_N(x)\no\\[2pt]
&&\hspace{1.9cm}+\frac{\sh\eta\,\ch(\al_+\!+\!\be_+\!-\!2(\tM\!-\!x)\eta)}{\sh(\al_+)\ch(\be_+)}\tilde\phi_N(x).\label{Right;right}
\eee
\end{widetext}
We note the useful identities
\begin{align}
&\phi_n(x)\,\sigma_n^z=\pm\frac{\ch\eta}{\sh\eta}\phi_n(x)\mp\frac{\eE^{\mp\eta}}{\sh\eta}\phi_n(x\pm1),\label{psi;phi}\\
&\sigma_n^z\,\tilde\phi_n(x)=\pm\frac{\ch\eta}{\sh\eta}\tilde\phi_n(x)\mp\frac{\eE^{\mp\eta}}{\sh\eta}\tilde\phi_n(x\pm1).\label{psi;phi2}
\end{align}
Using Eqs.~(\ref{bulk_1})-(\ref{right})  and (\ref{psi;phi}) repeatedly, we get Eqs.~(\ref{H00})-(\ref{HNN}).
Some identities, used in our calculations, are:
\bee
&&E_0=-w_--w_++4\De,\\
&&\frac{\sh\eta\,\ch(\al_\pm\!+\!\be_\pm)}{\sh(\al_\pm)\ch(\be_\pm)}=w_\pm-2\De,\label{Parameterization;1}
\eee
\bee
&&\frac{\sh\eta\,\ch(\al_\pm\!+\!\be_\pm\!+\!\eta)}{\sh(\al_\pm)\ch(\be_\pm)}=a_\pm b_\pm-1,\label{Parameterization;2}\\
&&\frac{\sh\eta\,\ch(\al_\pm\!+\!\be_\pm\!+\!2\eta)}{\sh(\al_\pm)\ch(\be_\pm)}=2a_\pm b_\pm\De-w_\pm.\label{Parameterization;3}
\eee

\section{Appendix D: The proof of Eqs.~(\ref{f;nm})-(\ref{f;NN}) }
\label{Proof}
Define the auxiliary function
\bee
W_{n,m}=\sum_{\sigma_1,\sigma_2=\pm}\left(A^{1,2}_{\sigma_1,\sigma_2}\,\eE^{\ir \sigma_1n \rP_1+\ir \sigma_2m\rP_2}\right.\no\\
\left.+{A}^{2,1}_{\sigma_2,\sigma_1}\,\eE^{\ir \sigma_2n \rP_2+\ir \sigma_1m\rP_1}\right),\label{F}
\eee
where $n,m$ are arbitrary integers.
Using BAE (\ref{BAE;M2}), the scattering matrix in (\ref{S_matrix}) and reflection matrices in (\ref{RM_1}), (\ref{RM_2}), one can get the following properties of $W_{n,m}$
\begin{align}
&\Lambda\, W_{n,m}=\sum_{\sigma=\pm 1}\left(W_{n+\sigma,m}+W_{n,m+\sigma}\right),\label{F;eq1}\\[2pt]
&2\De W_{n,n+1}=W_{n+1,n+1}+W_{n,n},\label{F;eq2}\\[2pt]
&w_-W_{0,n}=a_-b_-W_{1,n}+W_{-1,n},\label{F;eq3}\\[2pt]
&w_+W_{n,N}=a_+b_+W_{n,N-1}+W_{n,N+1},\label{F;eq4}\\[2pt]
&(\Lambda\,-2\De a_-b_--2\De)W_{0,0}\no\\
&=2\De(w_--2\De a_-b_-)W_{0,1},\label{F;eq5}\\[2pt]
&(\Lambda\,-2\De a_+b_+-2\De)W_{N,N}\no\\
&=2\De(w_+-2\De a_+b_+)W_{N-1,N}.\label{F;eq6}
\end{align}
With the help of Eqs.~(\ref{F;eq1})-(\ref{F;eq6}), we can prove that our
ansatz satisfies all the relations (\ref{f;nm})-(\ref{f;NN}). For
instance, Eq.~(\ref{f;1N}) can be proved as follows
\begin{align}
&\left(\Lambda\,-w_--w_+\right)f_{0,N}\no\\
&=g_{0,N}\left(\Lambda\,-w_--w_+\right)W_{0,N}\no\\
&=g_{0,N}\left[(1-a_-b_-)W_{1,N}+(1-a_+b_+)W_{0,N-1}\right]\no\\
&=f_{1,N}+f_{0,N-1}.
\end{align}

\section{Appendix E: Possibility of a further  partitioning of the invariant subspaces on special manifolds}

Let us consider the special case: $a_\pm b_\pm=1$.  Under this specific
condition, from Eq.~(\ref{H0N}) the SHS $\langle0,N|$ is an eigenstate of $H$
\bee \langle 0,N|H=\left(w_-+w_+-4\De\right)\langle 0,N|.  \eee
This SHS $\langle 0,N|$
corresponds to a special limiting case solution of BAE (\ref{BAE;M2}) with
$\rP_1=-\ir \ln(a_-)$, $\rP_2=-\ir \ln(a_+)$. In fact, both numerator and
denominator on the left hand side of (\ref{BAE;M2}) become zero, but the ratio
stays finite.

The bra vectors $\langle 0,n|,\,n=0,\dots,N$, form another sub-subspace as follows
\begin{align}
&\langle 0,0|H \!=\!\left(4\De-w_--w_+\right)\langle 0,0|+4\De\,(w_--2\De)\langle 0,1|,\no\\
&\langle 0,n|H\!=\!\left(w_--w_+-4\De\right)\langle 0,n| +2\langle0,n\!-\!1|\no\\
&\hspace{1.5cm}+2\langle0,n\!+\!1|,\qquad 1\leq n\leq N\!-\!1,\no\\
&\langle 0,N|H\!=\!\left(w_-+w_+-4\De\right)\langle 0,N|.\label{Reduce;relation}
\end{align}

The phantom Bethe states belonging to the above invariant sub-subspace have the
form $\langle \Psi_2|=\sum_{n=0}^N\langle 0,n|\mathsf
f_{0,n}$. 
Guided by Eq.~(\ref{Reduce;relation}) we propose $\mathsf
f_{0,n}$ to be a sum of plain waves.
\bee
\begin{aligned}\label{ansatz;Special;1}
	&
	{\mathsf f}_{0,n}=\mathsf A_+\eE^{\ir n\mathsf p}+\mathsf A_-\eE^{-\ir n\mathsf p},\quad n=1,\dots,N-1,\\
\end{aligned}
\eee
while $\mathsf f_{0,0},\,\mathsf f_{0,N}$ will be derived from the consistency conditions of (\ref{Reduce;relation}). Following Eqs. (\ref{Reduce;relation}) and (\ref{ansatz;Special;1}) we obtain
\begin{align}
\mathsf f_{0,0}=\frac{\mathsf A_++\mathsf A_-}{2\De\,(w_-\!-\!2\De)},\quad \mathsf f_{0,N}=\frac{\mathsf f_{0,N-1}}{2\cos(\mathsf p)-w_+},
\end{align}
and 
\bee
\frac{\sf A_-}{\sf A_+}=-\prod_{u=a_-,b_-}\frac{1\!-\!2\De\eE^{\ir\sf p}\!+\!u\eE^{\ir\sf p}}{1\!-\!2\De  u\!+\!u\eE^{\ir\sf p}}=-\eE^{2\ir N\mathsf p}.\label{Compatibility;Special}
\eee
The corresponding energy reads
$E=4\cos(\mathsf p)+w_--w_+-4\De$
where $\mathsf p$ satisfies the reduced BAE
\bee
\eE^{2\ir N\sf p}=\prod_{u=a_-,b_-}\frac{1\!-\!2\De\eE^{\ir\sf p}\!+\!u\eE^{\ir\sf p}}{1\!-\!2\De  u\!+\!u\eE^{\ir\sf p}}.\label{BAE;Reduce;1}
\eee

We can also get the same BAE (\ref{BAE;Reduce;1}) by letting $\rP_1, \rP_2$ in
the BAE (\ref{BAE;M2}) be $-\ir \ln (a_-)$ and $\sf p$ respectively ({note
	that for the Hermitian case the constants $a_{\pm},b_{\pm}$ are real}).
Noticing that $\pm\mathsf p$ are equivalent solutions and excluding two
trivial solutions $\mathsf p=0,\pi$, BAE (\ref{BAE;Reduce;1}) has $N$
independent non-trivial solutions.

Analogously, the bra vectors $\langle n,N|$, $n=0,\dots,N$, form another
sub-subspace. Suppose that $\langle\Psi_2|=\sum_{n=0}^N\langle n,N|\mathsf
f_{n,N}$. The coefficients $\{\mathsf f_{n,N}\}$ can be obtained via the
following transformation
\bee
\mathsf{f}_{n,N}\to \mathsf f_{0,N-n},\,\,\mbox{with}\,\, a_\pm,b_\pm,w_\pm\to a_\mp,b_\mp,w_\mp.\no
\eee
The corresponding energy is $E=4\cos(\mathsf p)+w_+-w_--4\De$ where the quasi-momentum $\mathsf p$ is a solution of the following BAE
\bee
\eE^{2\ir N\sf p}=\prod_{u=a_+,b_+}\frac{1\!-\!2\De\eE^{\ir\sf p}\!+\!u\eE^{\ir\sf p}}{1\!-\!2\De u\!+\!u\eE^{\ir\sf p}}.\label{BAE;Reduce;2}
\eee

The remaining $\binom{N}{2}-N$ eigenstates span the full $G_2^{+}$ basis. The two
reflection matrices  in (\ref{RM_1}) and (\ref{RM_2}) become  $-1$ and
$-\eE^{2\ir N\rP_k}$ respectively. In this case, the ``boundary terms" in
Eq.~(\ref{BAE;M2}) vanish and the BAE (\ref{BAE;M2}) acquire a simple form
\begin{align}
\eE^{2\ir N{\rP}_j}\!=\!\prod_{\sigma=\pm}\prod_{k\neq
	j}\frac{1-2\De\,\eE^{\ir{\rP}_j}+\eE^{\ir{\rP}_j+\ir
		\sigma{\rP}_k}}{1-2\De\,\eE^{\ir \sigma{\rP}_k}+\eE^{\ir{\rP}_j+\ir
		\sigma{\rP}_k}}, \,\, j\!=\!1,2.\label{BAE;Reduce;3}
\end{align} 

To sum up, in the special case we consider, the set of Bethe root pairs
$\{\rP_1,\rP_2\}$ in the original BAE (\ref{BAE;M2}) splits into $4$ subsets:
\begin{itemize}
	\setlength{\itemsep}{0pt}
	\setlength{\parsep}{0pt}
	\setlength{\parskip}{0pt}
	\item[(i)] one pair $\{\rP_1,\rP_2\}\!=\!\{ -\ir \ln (a_+), -\ir \ln (a_-)\}$ corresponding to SHS $\bra{0,N}$,
	\item[(ii)] $N$  pairs $\{\rP_1,\mathsf p\}$ with $\rP_1= -\ir\ln (a_-)$ 
	and $\mathsf p$ given by the solution of (\ref{BAE;Reduce;1}),
	\item[(iii)] $N$  pairs $\{\rP_1,\mathsf p\}$ with $\rP_1= -\ir\ln (a_+)$ 
	and $\mathsf p$ given by the solution of (\ref{BAE;Reduce;2}),
	\item[(iv)] $\binom{N}{2}-N$ pairs $\{\rP_1,\rP_2\}$ given by the solution of BAE (\ref{BAE;Reduce;3}).
\end{itemize}
In total, there are $1+N+ \binom{N}{2}=\dim \ G_2^{+}$ solutions, as expected.

Our example shows that there can be further partitionings of $G_2^{+}$, which
for $a_\pm b_\pm=1$ leads to three internal invariant subspaces of dimension $1,N,N$ within $G_2^{+}$ which are invariant w.r.t.~the action of $H$.
Likewise, if just one of the two conditions $a_\pm b_\pm=1$ is satisfied, some
internal invariant subspaces disappear, while others remain. Under other
constraints (arising when the coefficients of some ``unwanted'' terms on
the RHS of Eqs.~(\ref{H00})-(\ref{HNN}) vanish) various internal invariant
subspaces of $G_2^{+}$ can appear.

\bibliographystyle{IOP2}

\end{document}